\documentclass[preprint,proceedings]{rmaa}

\usepackage{paralist}

\usepackage{psfrag,color}

\SetYear{2002}
\SetConfTitle{Galaxy Evolution: Theory and Observations}

\title{Properties of Disk Galaxies in Phenomenological Models of Galaxy Formation} 

\author{
  C. D. Rimes\altaffilmark{1}
  and E. van Kampen\altaffilmark{1}
}

\altaffiltext{1}{Institute for Astronomy, University of Edinburgh, U.K.}

\shortauthor{Rimes \& van Kampen}
\shorttitle{Disk galaxy formation}

\fulladdresses{
\item Christopher D. Rimes and Eelco van Kampen:
  Institute for Astronomy, University of Edinburgh,
  Royal Observatory, Blackford Hill, Edinburgh EH9 3HJ, U.K.
  (cdr@roe.ac.uk, evk@roe.ac.uk)
}

\listofauthors{C. D. Rimes \& E. van Kampen}
\indexauthor{Rimes, C. D.}
\indexauthor{van Kampen, E.}

\abstract{We present a model for understanding the origins and evolution of galaxy morphologies from a phenomenological
perspective.  The model includes an observationally motivated prescription for star formation in galaxy disks, as well
as a merger-driven starburst mode of star formation.  We consider the formation of bulges and ellipticals both in
mergers and by global instabilities in disks.  We use our model to investigate the fundamental properties of disk
galaxies and the effects of surface-brightness limits on large galaxy surveys.
}

\addkeyword{Galaxies: formation}

\begin{document}
\maketitle

\section{Introduction}
\label{sec:intro}

Our understanding of galaxy formation in a cosmological context has increased considerably over the last decade through
the use of phenomenological models.  In these models, stars form primarily in galaxy disks and bulges are assumed to
form when two or more disk galaxies merge, or when bar instabilities lead to vertical heating of the disk.  We use such
a model for the evolution of galaxy morphologies to investigate the statistics of the fundamental parameters of disk
galaxies.  These results are particularly important for the interpretation of magnitude- or diameter-limited galaxy
surveys, which may miss a significant population of low-surface-brightness (LSB) galaxies.

\section{Model ingredients}
\label{sec:model}

The model used in this work is presented and discussed in detail in a paper soon to be published (Rimes, van Kampen \&
Peacock, in preparation).  The basic ingredients are:
\begin{list}{{\arabic{enumi}}.}{\usecounter{enumi}\leftmargin 2em \itemindent 0pt \labelsep 0.5em
\itemsep 0em \parsep 0em \topsep 0em \partopsep 0em}
\item the merging history of dark-matter haloes, taken directly from an N-body simulation;
\item galaxy-galaxy merging and hot gas stripping;
\item the formation and evolution of gas disks, including the gravitational effects of the baryons;
\item star formation in galaxy disks according to a Schmidt law ($\Sigma_{\rm SF} \propto \Sigma_{\rm g}^{1.4}$) with a
cut-off given by the Kennicutt threshold (Kennicutt 1989);
\item a merger-driven bursting mode of star formation;
\item injection of gas and metals from star formation into the cold ISM;
\item reheating of the cold ISM by supernovae;
\item stellar population synthesis, including the effects of dust extinction.
\end{list}

\section{Properties of disk galaxies at \lowercase{$z=0$}}
\label{sec:disks}

\begin{figure}[!t]
  \includegraphics[width=\columnwidth]{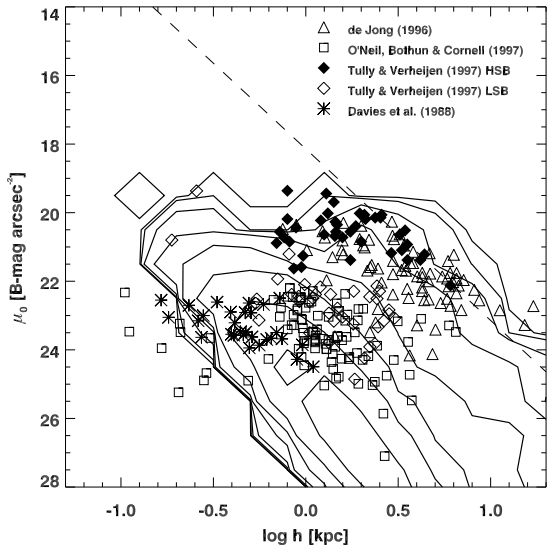}
  \caption{Bivariate distribution of central disk surface-brightness against exponential scale length.  Contours
  are the number density of galaxies as predicted by our model.  The dashed line is the knee of the Schechter
  luminosity function.
  \label{fig:1}}
\end{figure}

In figure \ref{fig:1} we plot the number density of galaxies as a function of central disk surface-brightness and
exponential scale length in the {\it B\/}-band, as predicted by our model.  The predicted distribution covers the
full range of the observational data, which include surveys of both high and low surface-brightness galaxies, and extends well
beyond the limits of current observations, with no significant fall-off in numbers seen until $\mu_0 \ga 30\;
B\mbox{\rm -mag\,as}^{-2}$.
The existence of large numbers of LSB galaxies down to $\mu_0 = 25\;B\mbox{\rm -mag\,as}^{-2}$ has been recognised for some time
(e.g. Impey \& Bothun 1997) and has important consequences both for observations and for theories of galaxy formation.

The lack of small, ultra-high surface-brightness galaxies is a consequence of disk
self-gravity.  Compact, self-gravitating disks are susceptible to bar instabilities (Efstathiou, Lake \& Negroponte
1982), which lead to vertical heating and the eventual destruction of the disk.

\section{Selection effects in galaxy surveys}
\label{sec:selection}

\begin{figure}[!t]
  \includegraphics[width=\columnwidth]{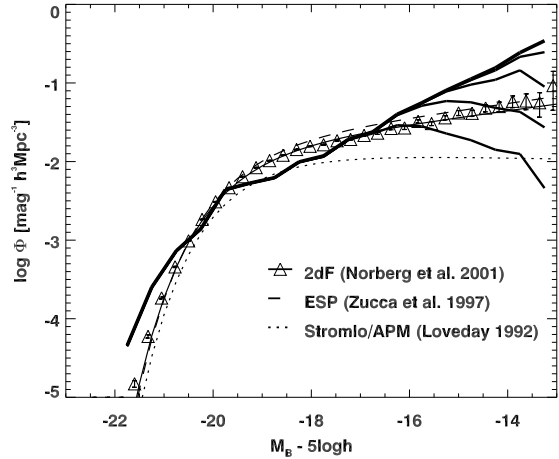}
  \caption{The effect of isophotal limits on the field galaxy luminosity function.  The heaviest solid line is the
  intrinsic luminosity function; the others are the predicted measurements for surveys with isophotal limits of (top
  to bottom) 26, 25, 24 and 23 $B\;\mbox{\rm -mag\,as}^{-2}$.
  \label{fig:2}}
\end{figure}

Understanding the processes that determine the spatial distribution of light in galaxies is important because it is
this, not the total flux, that primarily determines whether or not a galaxy is included in an observational sample.  It
has been pointed out (Cross et al. 2001) that a large part of the variation in measurements of the faint-end slope of
the luminosity function may be due to the different limiting isophotes of the surveys used.  To test this, we created
mock surveys of a $7\arcdeg \times 7\arcdeg$ area of sky by laying eight N-body simulation boxes end-to-end along the
line of sight and selecting galaxies with isophotal magnitudes of $B_{\rm iso}<25$ and isophotal diameters of $d_{\rm
iso}>2\arcsec$, for different limiting isophotes.  In figure \ref{fig:2} we compare the intrinsic {\it B\/}-band
luminosity function of the galaxies in our simulation with the luminosity functions derived from our mock surveys.
The intrinsic luminosity function has a faint-end slope far steeper than any of the observational measurements plotted
in figure \ref{fig:2}, a problem that has traditionally\adjustfinalcols been resolved by invoking strong feedback from
supernovae to suppress star formation in dwarf galaxies.  Our results demonstrate that, if selection effects in the observed samples
are accounted for, strong feedback may no longer be necessary (our fiducial model has weak feedback).

In a future paper, we intend to model selection effects in real galaxy surveys and show that this can resolve many
of the discrepancies between theory and observations.

%

\end{document}